\documentclass[11pt]{article}
\usepackage{natbib,epsfig,multirow,color,comment,ctable}
\usepackage{enumitem}
\usepackage[colorlinks=TRUE]{hyperref}
\usepackage{lscape} 

\usepackage{amsthm,amsmath,amssymb,colonequals,multirow,graphicx,epstopdf,marvosym,bm,hyperref}
\usepackage{caption}
\usepackage{graphicx}

\newcommand{\vecx}{\mathbf{x}}
\newcommand{\vecN}{\mathbf{N}}
\newcommand{\vecw}{\mathbf{w}}

\newcommand{\vecX}{\mathbf{X}}

\newcommand{\vectheta}{\mbox{\boldmath$\theta$}}

\newcommand{\varthet}{\mbox{\boldmath$\vartheta$}}
\newcommand{\vecmu}{\mbox{\boldmath$\mu$}}

\newcommand{\vecalpha}{\mbox{\boldmath$\alpha$}}

\newcommand{\mSigma}{\mbox{\boldmath$\Sigma$}}

\newcommand{\GIG}{\mathop{\rm GIG}}

\newcommand{\Tr}{\mathop{\rm Tr}}
\newcommand{\Lc}{\mathcal{L}}

\newcommand{\sumn}{\sum_{i = 1}^{N}}

\newcommand{\thetav}{\mbox{\boldmath{$\theta$}}}

\def\half{\frac{1}{2}}
\def\E{\mathbb E}
\def\Id{\mathbb I} 

\def\Vv{\mathbf V}

\def\av{\mathbf a}

\def\yv{\mathbf y}

\def\tv{\mathbf t}

\def\1v{\mathbf 1}
\def\0v{\mathbf 0}

\begin{document}

\title{Tackling the infinite likelihood problem when fitting mixtures of shifted asymmetric Laplace distributions}
\author{Yuan Fang\thanks{School of Pharmacy and Pharmaceutical Sciences, Binghamton University, State University of New York, 4400 Vestal Parkway East, Binghamton, NY 13902, USA. E-mail: yfang8@binghamton.edu},
~Brian C. Franczak\thanks{Department of Mathematics \& Statistics, MacEwan University, Edmonton, Alberta, T5J 4S2, Canada. E-mail: franczakb@macewan.ca}, and
Sanjeena Subedi\thanks{School of Mathematics \& Statistics, 4302 Herzberg Laboratories, Carleton University, 1125 Colonel By Drive, Ottawa, Ontario, K1S 5B6, Canada. E-mail: sanjeena.dang@carleton.ca}}
\date{\today}
\maketitle

\begin{abstract}
Mixtures of shifted asymmetric Laplace distributions were introduced as a tool for model-based clustering that allowed for the direct parameterization of skewness in addition to location and scale. Following common practices, an expectation-maximization algorithm was developed to fit these mixtures. However, adaptations to account for the `infinite likelihood problem' led to fits that gave good classification performance at the expense of parameter recovery. In this paper, we propose a more valuable solution to this problem by developing a novel Bayesian parameter estimation scheme for mixtures of shifted asymmetric Laplace distributions. Through simulation studies, we show that the proposed parameter estimation scheme gives better parameter estimates compared to the expectation-maximization based scheme. In addition, we also show that the classification performance is as good, and in some cases better, than the expectation-maximization based scheme. The performance of both schemes are also assessed using well-known real data sets.
\end{abstract}

\noindent\textbf{Keywords:} Finite mixture models; classification; statistical learning; shifted asymmetric Laplace distributions; expectation-maximization algorithm; Bayesian parameter estimation

\section{Introduction}\label{sec:intro}

The use of finite mixture models for classification has gained considerable attention in the associated
statistical literature \citep[see the reviews of][for thorough discussions on this topic]{McLBas88,McLPee00b,McN16,bouveyron19}. The density of a parametric finite mixture model can be written as
\begin{equation}\label{eq:fmm_dens}
f(\vecx\mid\bm\vartheta) = \sum_{g=1}^G{\pi_gf_g\left(\vecx\mid\vectheta_g\right)},
\end{equation}
where $\pi_g > 0$, such that $\sum_{g=1}^G{\pi_g} = 1$, are the mixing proportions, $f_1\left(\vecx\mid\vectheta_1\right),\ldots,f_G\left(\vecx\mid\vectheta_G\right)$ are the component density functions, and $\bm\vartheta = \left(\bm\pi,\vectheta_1,\ldots,\vectheta_G\right)$ is the vector of model parameters with $\bm\pi = \left(\pi_1,\ldots,\pi_G\right)$. At the time of \cite{fraley02a}, the typical practice was to assume that the component density functions belong to a multivariate Gaussian distribution. However, the use of mixtures of Gaussian distributions (MGDs) present a number limitations. For example, estimating the parameters of MGDs can be negatively impacted by observations that are far from the means of the assumed component Gaussian distributions and MGDs can not directly model skewness, i.e., they can not naturally account for asymmetry in the observed data. At the time of this paper, there have been a number of different models discussed in the literature which naturally address these limitations. For example, \cite{peel00} considered mixtures of multivariate-$t$ distributions (MMtDs), \cite{lin09}, \cite{cabral12}, and \cite{Lee14} discuss model fitting for mixtures of multivariate skew-normal distributions (MSNDs), \cite{lin10}, \cite{lee11}, and \cite{vrbik12} explore mixtures of multivariate skew-$t$ distributions (MStDs), \cite{lin14a} present the mixtures of multivariate skew $t$-normal distributions (MStNDs), \cite{franczak14} introduced mixtures of multivariate shifted asymmetric Laplace distributions (MSALDs), \cite{BroMcN15} developed mixtures of multivariate generalized hyperbolic distributions (MGHDs), and \cite{McNicholas17} proposed the mixtures of multivariate variance-gamma distributions (MVGDs). In each case, the proposed mixture models use either one, or a combination of, scaling and skewness parameters to address for the aforementioned limitations of the MGDs.

In this paper, we focus our attention on the MSALDs (note that a formal definition of this model is given in section~\ref{sec:mix_sal}). \cite{franczak14} present the MSALDs as a modest alternative to the MGDs that parameterize skewness in addition to location and scale. To estimate the parameters of the MSALDs, an expectation-maximization (EM) algorithm \citep{DemLai77} is developed. Notably, this EM algorithm is unique in that it is modified to account for the `infinite likelihood problem'. In spite of this modification, the MSALDs are still shown to provide an improved classification performance when compared to the MGDs in both a simulation study and real data analyses. However, a closer look at these results reveals that the utilized modification has a negative effect on parameter estimation. To address this issue, we propose a fully Bayesian parameter estimation scheme. 

For MGDs, Bayesian parameter estimation dates back at least as far as \cite{diebolt94} who developed a Gibbs sampling approach for estimating the parameters of finite mixture models with a fixed number of components. In \cite{richardson97} a literature review on Bayesian parameter estimation for finite mixture models when the number of components are either fixed or unknown is given. Further, the authors propose a reversible jump Markov Chain Monte Carlo (MCMC) method to estimate the parameters of finite mixture models with an unknown number of components. In \cite{stephens00b} the authors propose an alternative approach to the aforementioned reversible jump MCMC method. In \cite{stephens00a}, the authors give a solution to the `label-switching' problem that can lead to meaningless solutions that are not useful in classification problems. \cite{fraley02a} show that Bayesian estimation can mitigate the convergence issues that are found with EM based parameter estimation for MGDs. 

For mixtures of non-Gaussian distributions, \cite{fruhwirth06} provide a thorough review of Bayesian parameter estimation for a variety of finite mixture models. Among other distributions, the authors discuss mixtures of student-$t$, exponential, Poisson, and Binomial distributions. In \cite{fruhwirth10} and \cite{maleki19}, Bayesian parameter estimation schemes are introduced for MSNDs and MStDs, and mixtures of multivariate unrestricted skew-normal generalized hyperbolic distributions (MUSNGHDs), respectively. In \cite{hejblum19}, the authors discuss a novel Bayesian framework for clustering flow-cytometry data via Dirichlet process mixtures (DPM) and MStDs.

The remainder of this paper is organized as follows: in Section~\ref{sec:background} the relevant background materials are provided. In section~\ref{sec:methodology}, we present a novel Bayesian parameter estimation scheme for the MSALDs. In section~\ref{sec:gibbs_sampler}, we provide details on initialization and convergence of the proposed scheme and other computational considerations. In sections~\ref{sec:sim_study} and~\ref{sec:real_data}, we illustrate the effectiveness of our fully Bayesian parameter estimation in comparison to the EM algorithm of \cite{franczak14}. In section~\ref{sec:discuss}, we conclude with a discussion and suggestions for future work. 

\section{Background}\label{sec:background}

\subsection{Normal mean-variance mixtures}\label{sec:nmvm}

Suppose $w\in\mathbb{R}^+$ is a univariate random variable, $\vecmu\in\mathbb{R}^p$ is a location parameter, $\vecalpha\in\mathbb{R}^p$ is a skewness or drift parameter, and $\mSigma$ is a $p \times p$ covariance matrix. \cite{barndorff82} write that the distribution of a $p$-dimensional random variable $\vecX$ is a normal variance-mean mixture if $\vecX\mid W=w$ follows a multivariate Gaussian distribution with mean $\vecmu+w\vecalpha$ and covariance matrix $w\mSigma$. If $\vecalpha=0$ then a normal variance mixture is obtained. 

There are several well-known multivariate distributions that belong to the class of normal variance-mean mixtures. For example, the multivariate-$t$ distribution \citep{kotz04}, the generalized hyperbolic distribution \citep{mcneil05} and the asymmetric Laplace distribution \citep{kotz01} can all be expressed in terms of a multivariate Gaussian random variable (cf. Section~\ref{sec:em_alg}).

\subsection{Generalized inverse Gaussian distributions}\label{sec:gig_dist}

The first proposition of a generalized inverse Gaussian (GIG) distribution was in \cite{good53}. Herein, we take $Y\sim\text{GIG}\left(y\mid\phi,\chi,\nu\right)$ to mean that a random variable $Y\in\mathbb{R}^+$ follows a generalized inverse Gaussian distribution with $\phi,\chi\in\mathbb{R}^+$ and index $\nu\in\mathbb{Z}^+$. The density of $Y\sim\text{GIG}\left(y\mid\phi,\chi,\nu\right)$ can be written as
\begin{equation}\label{eq:gig_dens}
f_\text{GIG}(y\mid\phi,\chi,\nu) = \frac{(\phi/\chi)^{\nu/2}y^{\nu-1}}{2K_\nu(\sqrt{\phi\chi})}\exp{\left\{ -\frac{\phi y + \chi/y}{2} \right\}},
\end{equation}
for $y>0$, where $\phi,\chi~\text{and}~\nu$ are as previously defined and $K_\nu(\cdot)$ is the modified Bessel function of the third kind with index $\nu$. There are several special cases of the GIG distribution, such as the gamma distribution ($\chi=0$, $\nu>0$), the inverse Gaussian distribution ($\nu=-1/2$), the reciprocal gamma variate ($\phi=0$, $\nu<0$) and the reciprocal inverse Gaussian variate ($\nu=1/2$). 

The GIG distribution has been extensively studied within the literature. \cite{blaesild78} computed moments, cumulants and studied the shape of the density, \cite{halgreen79} investigated probabilistic properties, and \cite{barndorff77} came across the distribution while studying a representation of the hyperbolic distributions as a mixture of normal distributions. The statistical properties of the GIG distribution are given in \cite{jorgensen82}. Of specific interest are the tractability of the expected values of a GIG random variable. In particular, for $Y\sim\text{GIG}\left(y\mid\phi,\chi,\nu\right)$,
\begin{equation*}
\mathbb{E}\left[Y\right] = 
\sqrt{ \frac{\chi}{\phi} } \frac{ K_{\nu+1}\left(\sqrt{\phi\chi}\right) } { K_{\nu}\left(\sqrt{\phi\chi}\right) }
~\text{and}~
\mathbb{E}\left[1/Y\right] = 
\sqrt{ \frac{\phi}{\chi} } \frac{ K_{\nu+1}\left(\sqrt{\phi\chi}\right) } {
K_{\nu}\left(\sqrt{\phi\chi}\right) } - \frac{2\nu}{\chi}.
\end{equation*}

\subsection{Mixtures of shifted asymmetric Laplace distributions}\label{sec:mix_sal}

\cite{franczak14} introduce the MSALDs as a tool for classifying fully observed real multivariate observations. Formally, they write that $p$-dimensional random vector $\vecX$ follows a multivariate SAL distribution, i.e., $\vecX\sim\mathcal{SAL}_p(\vecmu,\vecalpha,\mSigma)$, if its density can be written as
\begin{equation}\label{eq:sal_dens}
f_\text{SAL}(\vecx \mid \vecmu,\vecalpha,\mSigma)=
\frac{2 \text{exp}\{(\vecx-\vecmu)'\mSigma^{-1}\vecalpha\}}{(2\pi)^{p/2}\lvert\mSigma\rvert^{1/2}}
\left(\frac{\delta(\vecx,\vecmu\mid\mSigma)}{2+\vecalpha'\mSigma^{-1}\vecalpha}  \right)^{\upsilon/2}K_{\upsilon}(u),
\end{equation}
where $u=\sqrt{(2+\vecalpha'\mSigma^{-1}\vecalpha)\delta(\vecx,\vecmu\mid\mSigma)},~\delta(\vecx,\vecmu\mid \mSigma) = \left(\vecx-\vecmu\right)'\mSigma^{-1}\left(\vecx-\vecmu\right)$ is the Mahalonobis distance between $\vecx$ and $\vecmu$, $\nu = (2 - p)/2$, $\vecmu, \vecalpha \in \mathbb{R}^p$ are the location and skewness parameters, respectively, and $\mSigma$ is a $p\times p$ covariance matrix \citep[cf.][]{kotz01}. It follows that the density of the MSALDs is found by replacing the component density functions in~\eqref{eq:fmm_dens} with the density given in~\eqref{eq:sal_dens}. This gives
\begin{equation}\label{eq:msal_dens}
f_\text{MSAL}(\vecx\mid\varthet)=\sum_{g=1}^{G} \pi_{g}\ f_\text{SAL}(\vecx\mid \vecmu_g,\vecalpha_g,\mSigma_g),
\end{equation}
where all terms are as previously defined for $g = 1,\ldots, G$.

\subsection{The expectation-maximization algorithm}\label{sec:em_alg}

The EM algorithm is a common choice for estimating the parameters of finite mixture models \citep[see][for examples]{McLKri08}. Formally, the EM algorithm is an iterative procedure that is used to find maximum likelihood estimates (MLEs) in the presence of missing, or unobserved, data. The computations of the EM algorithm are based on the complete-data, i.e., the union of the observed and unobserved data. On the expectation (E)-step, the expected value of the complete-data log-likelihood (CDLL) is computed. On the maximization (M)-step, the CDLL is maximized with respect to the model parameters.

For the MSALDs, the complete-data is made up of the unobserved group membership labels, denoted $z_{ig}$, where $z_{ig} = 1$ if observation $i$ is in group $g$ and is equal to 0 otherwise, and the latent variables $W_{ig}$, for $i=1,\ldots,n$ and $g=1,\ldots,G$. Since $\vecX\sim\mathcal{SAL}_p(\vecmu,\vecalpha,\mSigma)$ can be generated through the stochastic relationship $\vecX = \vecmu + W\vecalpha + \sqrt{W}\vecN$, where $\vecN\sim\mathcal{N}\left(\mathbf{0},\mSigma\right)$ and $W\sim\text{Exp}(1)$, we can show that 
\begin{equation}\label{eq:wgivenx}
W\mid\vecX=\vecx\sim\text{GIG}\left(\delta(\vecx,\vecmu\mid\mSigma),2+\vecalpha'\mSigma^{-1}\vecalpha,\nu\right). 
\end{equation}
Therefore, on the E-step of the EM algorithm for the MSALDs the latent $W_{ig}$ can be replaced by the expected values given in section~\ref{sec:gig_dist}. On the M-step of the EM algorithm for the MSALDs, the model parameters: $\pi_g$, $\vecalpha_g$, $\vecmu_g$, and $\mSigma_g$ are updated using their maximum likelihood estimators. Explicit details and an outline of this EM algorithm are given in section 3.2 of \cite{franczak14}. 

\subsection{The infinite likelihood problem}\label{sec:inf_like}

As the EM algorithm for the MSALDs iterates toward convergence, an issue can occur when updating $\vecmu_g$. When $p > 2$, the constant $\nu = (2-p)/2 $ is negative and the Malahanobis distance measure is now in the denominator of the density in \eqref{eq:sal_dens}. As a result, the updates for $\hat{\mu}_g$ can tend toward a value of $\vecx_i$. Unfortunately, while these values maximize the likelihood, they create computational issues when updating the remaining parameter values and the expected value of $1/W_{ig}$. Notably, the repercussions of these computational issues can also affect the results of bivariate applications.

To account for this issue, \cite{franczak14} propose a set-back approach where the update for $\vecmu_g$ is held fixed on the iteration before the absolute difference between any $\vecx_i$ and $\hat{\vecmu}_g$ falls below a user specified threshold. Using this fixed value of $\hat{\vecmu}_g$, $\hat{\vecalpha}_g$ is updated using a conditional MLE and $\hat{\mSigma}_g$ is updated in the normal way.

\section{Gibbs sampling}\label{sec:methodology}

We propose a Bayesian parameter estimation scheme based on Gibbs sampling for the MSALDs. Gibbs sampling is a well-known MCMC algorithm that generates samples from the posterior distributions of the parameters of interest. It is particularly effective when conjugate priors are employed and the conditional posterior distributions are in closed-form.

\subsection{The complete-data log-likelihood}

Herein, we take the marginal distribution of $\vecX$ to be a SAL distribution with density given in~\eqref{eq:sal_dens}. We can write the joint probability density of $\vecX\sim\mathcal{SAL}_p(\vecmu,\vecalpha,\mSigma)$ and $W\sim\text{Exp}(1)$ as
\begin{equation*}
\begin{split}
f(\vecx,w) &= f(\vecx\mid w)f(w)\\
&= (2\pi)^{-1/2}|w\mSigma|^{-1/2}\exp\left\{-\half (\vecx - \vecmu - w\vecalpha)'(w\mSigma)^{-1}(\vecx - \vecmu - w\vecalpha) \right\} \times \exp\{-w\},
\end{split}
\end{equation*}
where all terms are as previously defined.

Given $n$ independent observations $\vecx = (\vecx_1,\dots,\vecx_n)$ from a $G$-component mixture of SAL distributions, the complete-data likelihood can be written in the exponential family format as follows:
\begin{align*}
\Lc_{c}(\boldsymbol{\theta}_1,\dots,\boldsymbol{\theta}_G) 
&= \prod_{g = 1}^{G}\prod_{i=1}^{n}[\pi_g f(\vecx_i,w_{ig}\mid\vecalpha_g, \vecmu_g, \mSigma_g)]^{z_{ig}} \\
&= \prod_{g = 1}^{G}\left\{[\pi_g r(\boldsymbol{\theta}_g)]^{t_{0g}}\cdot\prod_{i = 1}^{n}[h(\vecx_i,w_{ig})]^{z_{ig}}\times\exp \Tr\left\{\sum_{j = 1}^{5}
\phi_{jg}(\boldsymbol{\theta}_g)t_{jg}(\vecx,\vecw_{g})\right\}\right\},
\end{align*}
where the observed $\vecx_i$, latent $w_{ig}$, and missing $z_{ig}$ are as previously defined, $\boldsymbol{\theta}_g = (\vecalpha_g, \vecmu_g, \mSigma_g)$ denote the parameters related to the $g^{th}$ mixture component, $r(\thetav_g) = \lvert\mSigma_g\rvert^{-1/2}\exp\left\{-\vecmu_g'\mSigma_g^{-1}\vecalpha_g\right\}$, $h(\vecx_i,w_{ig})= w_{ig}^{-d/2}$, $t_{0g} = \sumn{z_{ig}}$,
and the component-specific functions for the parameters, $\phi_{jg}(\thetav_g)$, and sufficient statistics $\tv_{jg}(\vecx,\vecw_{g})$, for $j = 1,\dots,5$, are given by:
\begin{align*}
\phi_{1g} &= \vecalpha_g'\mSigma_g^{-1}, &\tv_{1g}(\vecx,\vecw_{g}) &= \sumn z_{ig}\vecx_i;\\
\phi_{2g} &= \vecmu_g'\mSigma_g^{-1}, &\tv_{2g}(\vecx,\vecw_{g}) &= \sumn \frac{z_{ig}}{w_{ig}}\vecx_{i};\\
\phi_{3g} &= -(\mSigma_g^{-1}\vecalpha_g\vecalpha_g'+2\Id_{d}), &t_{3g}(\vecx,\vecw_{g}) &= \half \sumn z_{ig}w_{ig};\\
\phi_{4g} &= -(\mSigma_g^{-1}\vecmu_g\vecmu_g'), &t_{4g}(\vecx,\vecw_{g}) &= \half \sumn \frac{z_{ig}}{w_{ig}};\\
\phi_{5g} &= -\mSigma_g^{-1}, &\tv_{5g}(\yv,\vecw_{g}) &= \half \sumn \frac{z_{ig}}{w_{ig}}\vecx_i\vecx_i';
\end{align*}
where $\Id_{d}$ is the $d\times d$ identity matrix. For convenience, herein we will refer to the  $\tv_{jg}(\vecx,\vecw_{g})$ as $\tv_{jg}$, for $j=1,\ldots,5$ and $g=1,\ldots,G$.

\subsection{Prior and posterior distributions}\label{sec:prior_posterior}

Recall the definition of the conditional random variable $W\mid\vecX=\vecx$ given in \eqref{eq:wgivenx}. For the parameters $\thetav_g = (\vecalpha_g,\vecmu_g,\mSigma_g)$, we consider conjugate priors with hyperparameters $a_{0,g}^{(0)}$, $\av_{1,g}^{(0)}$, $\av_{2,g}^{(0)}$, $a_{3,g}^{(0)}$, $a_{4,g}^{(0)}$, and $\av_{5,g}^{(0)}$, where
\begin{equation*}
\av_{j,g} = \av_{j,g}^{(0)}+\tv_{jg},~\text{for}~j = 0,\ldots,5 
\end{equation*}
are the hyperparameters of the posterior distribution.

A conjugate Dirichlet prior distribution with hyperparameters $a_{0,1}^{(0)},\dots,a_{0,G}^{(0)}$ is assigned to the mixing proportions $\pi_1,\dots,\pi_G$. The resulting posterior is a Dirichlet distribution with hyperparameters $(a_{0,1},\dots,a_{0,G})$, i.e., $\text{Dir}(a_{1,0},\dots,a_{G,0})$.

A conjugate inverse Wishart prior is given to $\mSigma_g$, i.e. 
\begin{equation*}
\mSigma_g^{-1}\sim \text{Wishart}\left(a_{0,g}^{(0)},{\av_{5,g}^{(0)}}^{-1}\right),~\text{for}~g=1,\ldots,G;
\end{equation*}
Conditional on $\mSigma_g^{-1}$, the pair $(\vecmu_g,\vecalpha_g)$ is assigned a conjugate $2p-$dimensional multivariate Gaussian prior, i.e.,
\begin{equation*}
\left.\begin{pmatrix} 
\vecmu_g\\
\vecalpha_g
\end{pmatrix}
\right\vert
\mSigma_g^{-1}\sim\mathcal{N}_{2p}\left(
\begin{pmatrix} 
\vecmu_0^{(0)}\\ 
\vecalpha_0^{(0)}
\end{pmatrix},
\begin{pmatrix}
\tau_\mu^{(0)}\mSigma_g^{-1} & \tau_{\mu\alpha}^{(0)}\mSigma_g^{-1}\\
\tau_{\mu\alpha}^{(0)}\mSigma_g^{-1} & \tau_\alpha^{(0)}\mSigma_g^{-1}
\end{pmatrix}
\right),
\end{equation*}
where
\begin{equation*}
\vecmu_0^{(0)} = \frac{a_{3,g}^{(0)}\av_{2,g}^{(0)} - a_{0,g}^{(0)}\av_{1,g}^{(0)}}{a_{3,g}^{(0)}a_{4,g}^{(0)}-{a_{0,g}^{(0)}}^2},~
\vecalpha_0^{(0)} = \frac{a_{4,g}^{(0)}\av_{1,g}^{(0)} - a_{0,g}^{(0)}\av_{2,g}^{(0)}}{a_{3,g}^{(0)}a_{4,g}^{(0)}-{a_{0,g}^{(0)}}^2},~
\tau_\mu^{(0)} = a_{4,g}^{(0)},~
\tau_{\mu\alpha}^{(0)} = a_{0,g}^{(0)},~\text{and}~
\tau_\alpha^{(0)} = a_{3,g}^{(0)}.
\end{equation*}
It follows that the posterior distribution for $\mSigma_g$, which is also conditional on the pair $(\vecmu_g,\vecalpha_g)$, is given by 
\begin{equation*}
\mSigma_g^{-1}\mid\vecmu_g,\vecalpha_g,\cdot \sim \text{Wishart}\left(a_{0,g},\Vv_{0,g}\right),
\end{equation*} 
and the posterior distribution of $(\vecmu_g,\vecalpha_g)$ conditional on $\mSigma_g$ is 
\begin{equation*}
\left.\begin{pmatrix} \vecmu_g\\\vecalpha_g\end{pmatrix}\right\vert\mSigma_g^{-1},\cdot \sim\mathcal{N}_{2p}\left(\begin{pmatrix} \vecmu_{0,g}\\ \vecalpha_{0,g}\end{pmatrix},\begin{pmatrix}
\tau_{\mu,g}\mSigma_g^{-1} & \tau_{\mu\alpha,g}\mSigma_g^{-1}\\
\tau_{\mu\alpha,g}\mSigma_g^{-1} & \tau_{\alpha,g}\mSigma_g^{-1}
\end{pmatrix}\right),
\end{equation*} 
where
\begin{align*}
\Vv_{0,g}^{-1} &=\av_{5,g}+\vecmu_0^{(0)}\tau_{\mu}^{(0)}{\vecmu_0^{(0)}}'+\vecmu_0^{(0)}\tau_{\mu\alpha}^{(0)}{\vecalpha_0^{(0)}}'+\vecalpha_0^{(0)}\tau_{\mu\alpha}^{(0)}{\vecmu_0^{(0)}}'+\vecalpha_0^{(0)}\tau_{\alpha}^{(0)}{\vecalpha_0^{(0)}}'\\
&\hspace{1in}-\left(\vecmu_{0,g}\tau_{\mu,g}\vecmu_{0,g}'+\vecmu_{0,g}\tau_{\mu\alpha,g}\vecalpha_{0,g}'+\vecalpha_{0,g}\tau_{\mu\alpha,g}\vecmu_{0,g}'+\vecalpha_{0,g}\tau_{\alpha,g}\vecalpha_{0,g}'\right),
\end{align*}
with 
\begin{equation*}
\vecmu_{0,g} = \dfrac{a_{3,g}\av_{2,g} - a_{0,g}\av_{1,g}}{a_{3,g}a_{4,g}-{a_{0,g}}^2},~ \vecalpha_{0,g} = \dfrac{a_{4,g}\av_{1,g} - a_{0,g}\av_{2,g}}{a_{3,g}a_{4,g}-{a_{0,g}}^2},~
\tau_{\mu,g} = a_{4,g},~
\tau_{\mu\alpha,g} = a_{0,g},~\text{and}~
\tau_{\alpha,g} = a_{3,g}.
\end{equation*}
Given an estimate of $(\hat\vecalpha_g,\hat\vecmu_g,\hat\mSigma_g)$, the probability that $z_{ig} = 1$ is
\begin{equation}\label{eq:zig_up}
\hat{z}_{ig} = \frac{\hat\pi_g f_g(\vecx_i|w_{ig};\hat\vecalpha_g, \hat\vecmu_g, \hat\mSigma_g)\cdot f_g(w_{ig})}{\sum_{k = 1}^{G}{\hat\pi_k f_k(\vecx_i|w_{ik};\hat\vecalpha_k,\hat\vecmu_k,\hat\mSigma_k)\cdot f_k(u_{ik})}}.
\end{equation}

\section{Parameter estimation and other considerations}\label{sec:gibbs_sampler}
\subsection{Summary of the proposed Gibbs sampling framework}
The steps (denoted S0, \ldots, S3) of the proposed Gibbs sampling framework can be summarized as follows:
\begin{enumerate} 
	\item[S0] Initialization: For the observed data $\vecx = (\vecx_1,\dots,\vecx_n)$, the algorithm is initialized with $G$ components. 
	The $\hat z_{ig}$ are initialized using the results from $k$-means clustering.
	Given the $\hat z_{ig}$, the model parameters for the $g$th component are initialized as follows:
	\begin{enumerate}[label=\arabic*.]
		\item $\vecalpha_g$ is assigned a $d$-dimensional vector with all entries equal to $0.05$.
		\item $\vecmu_g$ is set as the component sample mean.
		\item $\mSigma_g$ is initialized as the component sample variance
		matrix.
		\item $\pi_g$ is set as the proportion of observation in the $g^{th}$ component.
	\end{enumerate}
	\item[S1] At $t^{th}$ iteration, update $z_{ig}$ using~\eqref{eq:zig_up}, where $(\hat\vecalpha_g,\hat\vecmu_g,\hat\mSigma_g)^{(t-1)}$ are the values of $\vecalpha_g,\vecmu_g$, and $\mSigma_g$ sampled from the $(t-1)^{th}$ iteration.
	\item[S2]
	\begin{enumerate}[label=\arabic*.]
		\item Update $W_i$ by drawing samples from a $\GIG\left(\delta(\vecx,\vecmu\mid\mSigma),2+\vecalpha'\mSigma^{-1}\vecalpha,\nu\right)$ distribution for $i = 1,\dots, n$;
		\item Based on the updated $W_{ig}$ and $z_{ig}$, update the hyperparameters \\ $\{a_{0,g},\av_{1,g},\av_{2,g},a_{3,g},a_{4,g},\av_{5,g}\}$, for $g = 1,\dots, G$;
		\item Update the parameters to  $(\vecalpha_g,\vecmu_g,\mSigma_g)^{(t)}$ by each drawing one sample from their posterior distributions described in Section~\ref{sec:prior_posterior};
		\item Compute $\delta(\vecx_i,\vecmu_g^{(t)}\mid\mSigma_g^{(t)})$ for all observations $i=1,\dots,n$. If it is smaller than $10^{-6}$ for any $i$, generate a new sample of $(\vecalpha_g,\vecmu_g,\mSigma_g)^{(t)}$.
		\item Update the mixing proportions to $(\pi_1^{(t)},\dots,\pi_G^{(t)})$ by drawing a sample their the posterior distribution, $\text{Dir}(a_{1,0},\dots,a_{G,0})$.
	\end{enumerate}
	\item[S3] Repeat S1 and S2 until convergence.
\end{enumerate}


\subsection{Convergence assessment and label switching}
To diagnose the convergence of Monte Carlo Markov Chains, three independent sequences, with different $k$-means initializations are simulated. The likelihood values are calculated using the updated parameters at the end of each iteration and the chains of likelihood values are monitored for convergence using the potential scale reduction factor \citep{gelman92}. After dropping values from the ``burn-in''period, if the potential scale reduction factor is below 1.1, the chains are considered to have converged and mixed well. We further draw 500 samples from all three chains, after reaching a stationary approximation of the posterior distribution; then following the suggestion of \cite{diebolt94}, we use averages of these samples to compute parameter estimates. To deal with the label switching issue, we follow the suggestion of \cite{richardson97} and put an artificial constraint on mixing proportions.

\subsection{Model selection and classification performance}


In our applications, we will compare the performance of Bayesian information criterion \citep[BIC;][]{schwarz78} and the integrated complete likelihood \citep[ICL;][]{biernacki00} when selecting the number of components for the MSALDs. The BIC is a popular tool for determining the number of components when a finite mixture model is used for classification. The BIC is given by:
\begin{equation}\label{eq:BIC}
\text{BIC}=2l(\vecx \mid \hat{\varthet}) - \rho\log n,
\end{equation}
where $l(\vecx\mid\hat{\varthet})$ is the maximized value of the log-likelihood, $\rho = G - 1 + 2Gp + Gp(p-1)/2$ is the number of free parameters, and $n$ is the number of observations in the dataset. For more details on the use of BIC in this context see \cite{CamFra97} and \cite{DasRaf98}. The ICL penalizes the BIC by subtracting a measure of the estimated entropy which represents classification uncertainity. The ICL is given by
\begin{equation}
\text{ICL}=\text{BIC}+\sum^n_{i=1}\sum^G_{g=1}\text{MAP}(z_{ig})\text{log}(z_{ig}),
\end{equation}
where $z_{ig}$ is the expected value of $Z_{ig}$, MAP($z_{ig}$) is the maximum \textit{a posteriori} classification given by the $z_{ig}$, and the BIC is as defined in \eqref{eq:BIC}.

To evaluate the classification performance of the fitted MSALDs we use the adjusted Rand index \citep[ARI;][]{HubAra85}. The ARI corrects the Rand index \citep[RI;][]{rand71} for chance, has an expected value equal to zero under random classification, and is equal to one when there is perfect class agreement. \cite{steinley04} provide a set of guidelines for interpreting the ARI. 


\section{Simulation studies}\label{sec:sim_study}


Herein, we use MSALD-Bayes and MSALD-EM to designate MSALDs fitted using the proposed Bayesian parameter scheme and the EM algorithm of \cite{franczak14}, respectively. We compare the performance of the MSALD-Bayes and the MSALD-EM using four simulation studies. For each simulation study we use the \texttt{rmsal} function from the {\sf R} package \texttt{MixSAL} to simulate 100 data sets from the MSALDs of interest. The MSALD-EM is implemented using the \texttt{msal} function in \texttt{MixSAL}. Like the MSALD-Bayes, the MSALD-EM is initialized using a $k$-means solution. If the algorithm fails to converge from the $k$-means solution then random partitions are  used until convergence is reached. For the MSALD-EM, convergence is determined using the stopping criterion given in \cite{Lin95} with $\epsilon = 0.01$. For each simulation study, we compare the parameter estimates returned by the MSALD-Bayes and MSALD-EM to the true parameter values. We conclude this section with a comparison of the classification performance, model selection criteria, and the average elapsed times to reach convergence for both schemes across all five simulation studies.

\subsection{Simulation study 1: two well separated SAL clusters}

In the first simulation study, we consider the example of two well separated SAL clusters given in \cite{franczak14}. Table~\ref{tab:sim1_pe} gives the true parameter values and the average of the estimates, with standard deviation, returned by the MSALD-Bayes and MSALD-EM across the 100 data sets simulated for this study. At a glance, it appears that both schemes return very good estimates of the true parameter values. 

\begin{table}[!ht]
\centering
\caption{True parameter values and the mean estimates with standard deviations returned by the MSALD-Bayes and MSALD-EM across the 100 simulated data sets generated for simulation 1.}
\label{tab:sim1_pe}
\begin{tabular}{cccc}
\toprule  
\textbf{Parameter} & \textbf{True values} & \textbf{MSALD-Bayes} & \textbf{MSALD-EM} \\
\midrule  
${\bm{\alpha}}_1$
&
$\left(
\begin{array}{c}
2 \\
2 \\
\end{array} \right)
$
&
\textcolor{black}{$
\left(
\begin{array}{c}
2.02 \pm 0.15 \\
2.01 \pm 0.16 \\
\end{array} 
\right)
$}
& 
$
\left(
\begin{array}{c}
1.93 \pm 0.15 \\
1.93 \pm 0.17 \\
\end{array} 
\right)
$
\\
\\
${\bm{\Sigma}}_1$
&
$
\left(
\begin{array}{cc}
1 & 0.5 \\
0.5 & 1 \\
\end{array} \right)
$
&
\textcolor{black}{$
\left(\begin{array}{cc}  
0.95 \pm 0.19 & 0.45 \pm 0.16 \\
0.45 \pm 0.16 & 0.95 \pm 0.18
\end{array}\right)
$}
&
$
\left(\begin{array}{cc}  
1.16 \pm 0.26 & 0.65 \pm 0.25 \\
0.65 \pm 0.25 & 1.14 \pm 0.28
\end{array}\right)
$
\\
\\
${\bm{\mu}}_1$
& 
$
\left(
\begin{array}{c}
0 \\
5 \\
\end{array} \right)
$
&
\textcolor{black}{$
\left(
\begin{array}{c}
-0.016 \pm 0.055 \\
4.988 \pm 0.055 
\end{array} 
\right)
$}
&
$
\left(
\begin{array}{c}
0.08 \pm 0.10 \\
5.08 \pm 0.10 
\end{array} 
\right)
$ \\
\\
$\pi_1$ 
&
0.5
& \textcolor{black}{$0.50 \pm 0.02$}
& $0.50 \pm 0.02$ \\
\midrule
${\bm{\alpha}}_2$
&
$\left(
\begin{array}{c}
2 \\
1 \\
\end{array} \right)
$
&
\textcolor{black}{$
\left(
\begin{array}{c}
2.03 \pm 0.15 \\
1.01 \pm 0.10 \\
\end{array} 
\right)
$}
& 
$
\left(
\begin{array}{c}
1.91 \pm 0.24 \\
0.94 \pm 0.13 \\
\end{array} 
\right)
$
\\
\\
${\bm{\Sigma}}_2$
&
$
\left(
\begin{array}{cc}
1 & 0 \\
0 & 1 \\
\end{array} \right)
$
&
\textcolor{black}{$
\left(\begin{array}{cc}  
0.98 \pm 0.20 & -0.00 \pm 0.13 \\
-0.00 \pm 0.13 & 1.02 \pm 0.14
\end{array}\right)
$}
&
$
\left(\begin{array}{cc}  
1.22 \pm 0.48 & 0.13 \pm 0.24 \\
0.13 \pm 0.24 & 1.07 \pm 0.16
\end{array}\right)
$
\\
\\
${\bm{\mu}}_2$
& 
$
\left(
\begin{array}{c}
0 \\
-2 \\
\end{array} \right)
$
&
\textcolor{black}{$
\left(
\begin{array}{c}
-0.01 \pm 0.07 \\
-2.01 \pm 0.06 
\end{array} 
\right)
$}
&
$
\left(
\begin{array}{c}
0.11 \pm 0.20   \\
-1.94 \pm 0.11
\end{array} 
\right)
$ \\
\\
$\pi_2$ 
&
0.5
& \textcolor{black}{$0.50 \pm 0.02$}
& $0.50 \pm 0.02$ \\
\bottomrule
\end{tabular}
\end{table}

\begin{figure}[!ht]
\centering
\includegraphics[width=0.95\linewidth,height=0.375\textheight]{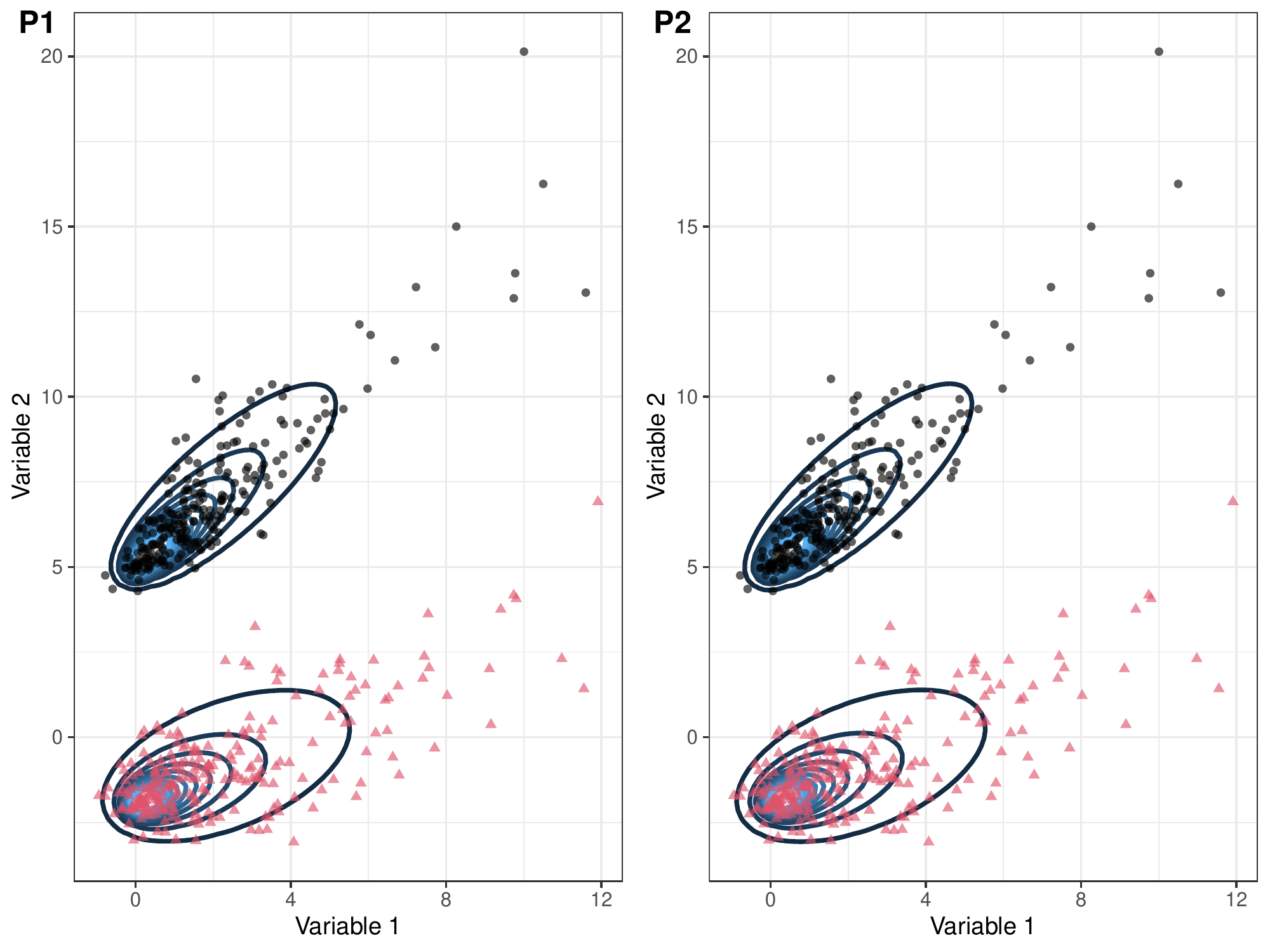}
\caption{Panel 1 (P1) and panel 2 (P2) give scatterplots showing the classification results with contours superimposed for the MSALD-Bayes and the MSALD-EM approaches, respectively, when fitted to the 1st data set generated for simulation study 1.}
\label{fig:sim1_contours}
\end{figure}

Panels 1 and 2 of Figure~\ref{fig:sim1_contours} display scatterplots of the classification solutions with fitted contour plots using the results from both schemes for the 1st data set simulated in simulation study 1. The contour plots imply that the fits from each scheme are nearly identical. Upon closer inspection, we can see that the average of values returned by MSALD-Bayes are closer to the true values than those returned by the MSALD-EM. In particular, Table~\ref{tab:sim1_pe} shows that the skewness parameter is consistently under-estimated by the MSALD-EM. This corresponds to an over estimation of the entries in the scale matrices. Further, the results in Table~\ref{tab:sim1_pe} also indicate that the MSALD-Bayes tends to return more precise estimates compared to the MSALD-EM as the standard deviations associated with the scale and location parameters are consistently higher for the MSALD-EM. 
 
\subsection{Simulation study 2: two slightly overlapping SAL clusters}

In the second simulation study, we move the SAL clusters from the first simulation study closer together to create a slightly more difficult classification problem. Table~\ref{tab:sim2_pe} gives the true parameter values and the average estimates, with standard deviations, returned by MSALD-Bayes and MSALD-EM. 

\begin{table}[!ht]
\centering
\caption{True parameter values and the mean estimates with standard deviations returned by the MSALD-Bayes and MSALD-EM across the 100 simulated data sets generated for simulation 2.}
\label{tab:sim2_pe}
\begin{tabular}{cccc}
\toprule  
\textbf{Parameter} & \textbf{True values} & \textbf{MSALD-Bayes} & \textbf{MSALD-EM} \\
\midrule  
${\bm{\alpha}}_1$
&
$\left(
\begin{array}{c}
2 \\
2 \\
\end{array} \right)
$
&
\textcolor{black}{$
\left(
\begin{array}{c}
2.02 \pm 0.17 \\
2.04 \pm 0.16 \\
\end{array} 
\right)
$}
& 
$
\left(
\begin{array}{c}
1.71 \pm 0.84 \\
1.54 \pm 1.12 \\
\end{array} 
\right)
$
\\
\\
${\bm{\Sigma}}_1$
&
$
\left(
\begin{array}{cc}
1 & 0.5 \\
0.5 & 1 \\
\end{array} \right)
$
&
\textcolor{black}{$
\left(\begin{array}{cc}  
0.98 \pm 0.20 & 0.45 \pm 0.18 \\
0.45 \pm 0.18 & 0.96 \pm 0.26 
\end{array}\right)
$}
&
$
\left(\begin{array}{cc}  
1.41 \pm 0.21 & 1.00 \pm 1.25 \\
1.00 \pm 1.25 & 1.88 \pm 1.94
\end{array}\right)
$
\\
\\
${\bm{\mu}}_1$
& 
$
\left(
\begin{array}{c}
0 \\
3 \\
\end{array} \right)
$
&
\textcolor{black}{$
\left(
\begin{array}{c}
-0.01 \pm 0.06 \\
2.99 \pm 0.07 
\end{array} 
\right)
$}
&
$
\left(
\begin{array}{c}
0.26 \pm 0.72 \\
3.29 \pm 0.77 \\
\end{array} 
\right)
$ \\
\\
$\pi_1$ 
&
0.5
& \textcolor{black}{$0.50 \pm 0.03$}
& $0.52 \pm 0.09$ \\
\midrule
${\bm{\alpha}}_2$
&
$\left(
\begin{array}{c}
2 \\
1 \\
\end{array} \right)
$
&
\textcolor{black}{$
\left(
\begin{array}{c}
1.99 \pm 0.18 \\
1.02 \pm 0.12
\end{array} 
\right)
$}
& 
$
\left(
\begin{array}{c}
1.89 \pm 0.25 \\
0.88 \pm 0.64
\end{array} 
\right)
$
\\
\\
${\bm{\Sigma}}_2$
&
$
\left(
\begin{array}{cc}
1 & 0 \\
0 & 1
\end{array} \right)
$
&
\textcolor{black}{$
\left(\begin{array}{cc}  
0.94 \pm 0.22 & -0.02 \pm 0.12 \\
-0.02 \pm 0.12 & 0.96 \pm 0.13
\end{array}\right)
$}
&
$
\left(\begin{array}{cc}  
1.09 \pm 0.32 & 0.12 \pm 0.40 \\
0.12 \pm 0.40 & 1.41 \pm 1.55
\end{array}\right)
$
\\
\\
${\bm{\mu}}_2$
& 
$
\left(
\begin{array}{c}
0 \\
-1
\end{array} \right)
$
&
\textcolor{black}{$
\left(
\begin{array}{c}
-0.02 \pm 0.06 \\
-1.01 \pm 0.05 
\end{array} 
\right)
$}
&
$
\left(
\begin{array}{c}
0.07 \pm 0.10  \\
-0.72 \pm 1.01
\end{array} 
\right)
$ \\
\\
$\pi_2$ 
&
0.5
& \textcolor{black}{$0.50 \pm 0.03$}
& $0.52 \pm 0.09$ \\
\bottomrule
\end{tabular}
\end{table}

The results in Table~\ref{tab:sim2_pe} highlight the improvement in parameter recovery found for the Bayesian estimate scheme. In particular, we notice that the estimates of the skewness parameters are once again under estimated by the MSALD-EM and that the associated standard deviations are much higher for all model parameters compared to the MSALD-Bayes. Looking at the scale matrices, we see that the average of the estimates for the off-diagonal elements for the first mixture component are very high compared to the true values and the average value returned by the MSALD-Bayes. These results imply that an increase in the difficulty of the classification problem has a negative effect on the EM-based estimation procedure. 

\subsection{Simulation study 3: the inverted trillium}

\begin{figure}[!ht]
\centering
\includegraphics[width=0.95\linewidth,height=0.375\textheight]{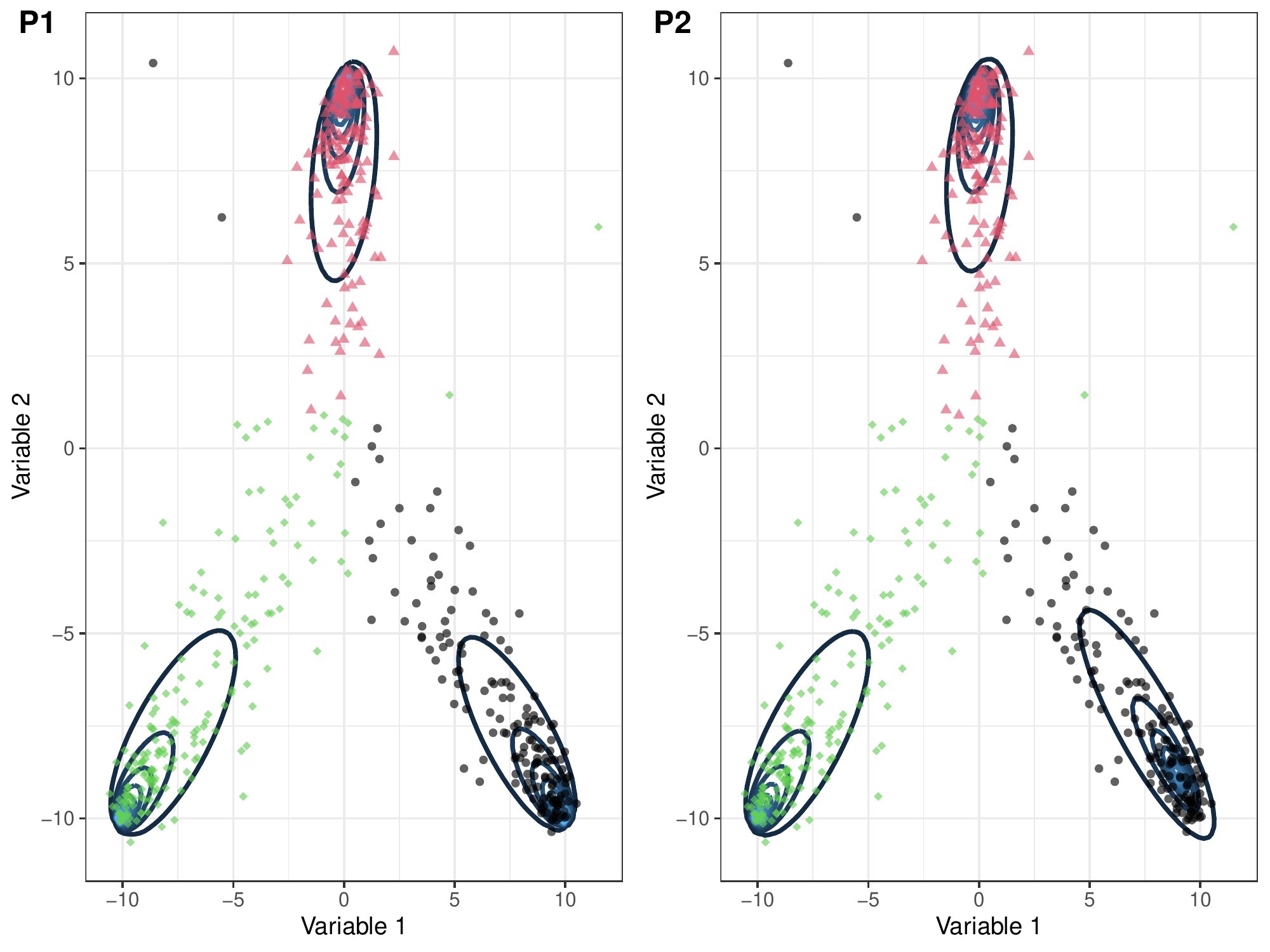}
\caption{Panel 1 (P1) and panel 2 (P2) give scatterplots showing the classification results with contours superimposed for the MSALD-Bayes and the MSALD-EM, respectively, when fitted to the 1st data set generated for simulation study 3.}
\label{fig:sim3_contours}
\end{figure}

In simulation study 3, we consider the inverted trillium problem which is created using three SAL clusters. Illustrations of a data set generated from this simulation study, with the classification solutions and fitted contours from the MSALD-Bayes (P1) and MSALD-EM (P2) superimposed, are presented in Figure~\ref{fig:sim3_contours}. Figure~\ref{fig:sim3_contours} shows that classification performance of the two approaches is nearly identical with only one disagreement, but there are noticeable differences in the fits. In particular, the tail behaviour between the two approaches is considerably different for group 1 (black dots) and group 2 (red triangles).

Table~\ref{tab:sim3_pe} gives the true parameter values and the average of the estimates, with standard deviations, returned by MSALD-Bayes and MSALD-EM. These results reinforce some of the features that were displayed in Figure~\ref{fig:sim3_contours} for group 1, that the MSALD-Bayes is returning estimates of the skewness parameters that are larger in value than those returned by the MSALD-EM. However, contrary to what was displayed in~Figure~\ref{fig:sim3_contours} we see that the estimates of the skewness parameters from the MSALD-Bayes are larger than those returned by the MSALD-EM, for all three groups. Furthermore, we all find that the average value of the estimates returned by the MSALD-Bayes for $\vecalpha_1,\ldots,\vecalpha_3$ are also closer to the true values compared to those returned by the MSALD-EM. 

Compared to simulation studies 1 and 2, we can see that, once again, the majority of the estimates from the MSALD-Bayes have smaller standard deviations compared to those returned from the MSALD-EM, with the exception of those values given for group 1. Furthermore, it is clear that the majority of the parameter estimates from the MSALD-Bayes are also closer to the true values, with the exception being the estimates for the mixing proportions. 

\begin{table}[!ht]
\centering
\caption{True parameter values and the mean estimates with standard deviations returned by the MSALD-Bayes and MSALD-EM across the 100 simulated data sets generated for simulation 3.}
\label{tab:sim3_pe}
\begin{tabular}{cccc}
\toprule  
\textbf{Parameter} & \textbf{True values} & \textbf{MSALD-Bayes} & \textbf{MSALD-EM} \\
\midrule  
${\bm{\alpha}}_1$
&
$\left(
\begin{array}{c}
0 \\
-3 \\
\end{array} \right)
$
&
\textcolor{black}{$
\left(
\begin{array}{c}
-0.01 \pm 0.13 \\
-3.30 \pm 0.50 \\
\end{array} 
\right)
$}
& 
$
\left(
\begin{array}{c}
-0.00 \pm 0.13 \\
-2.78 \pm 0.44 \\
\end{array} 
\right)
$
\\
\\
${\bm{\Sigma}}_1$
&
$
\left(
\begin{array}{cc}
1 & 0.5 \\
0.5 & 1 \\
\end{array} \right)
$
&
\textcolor{black}{$
\left(\begin{array}{cc}  
1.21 \pm 0.33 & 0.53 \pm 0.26 \\
0.53 \pm 0.26 & 1.00 \pm 0.51
\end{array}\right)
$}
&
$
\left(\begin{array}{cc}  
1.01 \pm 0.20 & 0.48 \pm 0.21 \\
0.48 \pm 0.21 & 1.68 \pm 0.87
\end{array}\right)
$
\\
\\
${\bm{\mu}}_1$
& 
$
\left(
\begin{array}{c}
0 \\
10 \\
\end{array} \right)
$
&
\textcolor{black}{$
\left(
\begin{array}{c}
0.01 \pm 0.08 \\
10.02 \pm 0.09 
\end{array} 
\right)
$}
&
$
\left(
\begin{array}{c}
0.00 \pm 0.09 \\
9.77 \pm 0.24 \\
\end{array} 
\right)
$ \\
\\
$\pi_1$ 
&
$0.\bar{3}$
& \textcolor{black}{$0.35 \pm 0.03$}
& $0.33 \pm 0.03$ \\
\midrule
${\bm{\alpha}}_2$
&
$\left(
\begin{array}{c}
3 \\
3 \\
\end{array} \right)
$
&
\textcolor{black}{$
\left(
\begin{array}{c}
3.02 \pm 0.35 \\
3.05 \pm 0.34 \\
\end{array} 
\right)
$}
& 
$
\left(
\begin{array}{c}
2.85 \pm 0.37 \\
2.84 \pm 0.36
\end{array} 
\right)
$
\\
\\
${\bm{\Sigma}}_2$
&
$
\left(
\begin{array}{cc}
1 & 0 \\
0 & 1
\end{array} \right)
$
&
\textcolor{black}{$
\left(\begin{array}{cc}  
1.05\pm 0.47 & -0.21 \pm 0.22 \\
-0.21 \pm 0.22 & 0.91 \pm 0.35
\end{array}\right)
$}
&
$
\left(\begin{array}{cc}  
1.54 \pm 0.65 & 0.53 \pm 0.65 \\
0.53 \pm 0.65 & 1.53 \pm 0.68
\end{array}\right)
$
\\
\\
${\bm{\mu}}_2$
& 
$
\left(
\begin{array}{c}
-10 \\
-10
\end{array} \right)
$
&
\textcolor{black}{$
\left(
\begin{array}{c}
-10.01 \pm 0.08 \\
-10.03 \pm 0.07 
\end{array} 
\right)
$}
&
$
\left(
\begin{array}{c}
-9.82 \pm 0.19 \\
-9.82 \pm 0.20
\end{array} 
\right)
$ \\
\\
$\pi_2$ 
&
$0.\bar{3}$
& \textcolor{black}{$0.33 \pm 0.03$}
& $0.33 \pm 0.03$ \\
\midrule
${\bm{\alpha}}_3$
&
$\left(
\begin{array}{c}
-3 \\
3 \\
\end{array} \right)
$
&
\textcolor{black}{$
\left(
\begin{array}{c}
-2.96 \pm 0.30 \\
2.94 \pm 0.30
\end{array} 
\right)
$}
& 
$
\left(
\begin{array}{c}
-2.83 \pm 0.31 \\
2.80 \pm 0.32
\end{array} 
\right)
$
\\
\\
${\bm{\Sigma}}_3$
&
$
\left(
\begin{array}{cc}
1 & 0.25 \\
0.25 & 1
\end{array} \right)
$
&
\textcolor{black}{$
\left(\begin{array}{cc}  
0.94 \pm 0.31 & 0.33 \pm 0.16 \\
0.33 \pm 0.16 & 0.88 \pm 0.30
\end{array}\right)
$}
&
$
\left(\begin{array}{cc}  
1.55 \pm 0.66 & -0.29 \pm 0.60 \\
-0.29 \pm 0.60 & 1.50 \pm 0.65
\end{array}\right)
$
\\
\\
${\bm{\mu}}_3$
& 
$
\left(
\begin{array}{c}
10 \\
-10
\end{array} \right)
$
&
\textcolor{black}{$
\left(
\begin{array}{c}
10.04 \pm 0.07 \\
-10.02 \pm 0.07 
\end{array} 
\right)
$}
&
$
\left(
\begin{array}{c}
9.84 \pm 0.17 \\
-9.82 \pm 0.18
\end{array} 
\right)
$ \\
\\
$\pi_3$ 
&
$0.\bar{3}$
& \textcolor{black}{$0.33 \pm 0.03$ }
& $0.34 \pm 0.03$ \\
\bottomrule
\end{tabular}
\end{table}

\subsection{Simulation study 4: three-dimensional clusters}

In the fourth simulation study, we consider two three-dimensional SAL clusters. The skewness vectors for the 1st and 2nd component are generated from normal distributions with mean 2 and standard deviation 1 and mean 1 and standard deviation 2, respectively. For the 1st component, we set the mean vector equal to the zero vector and generate the entries of the mean vector for the 2nd component from a normal distribution with mean 0 and standard deviation 4. Both covariance matricies are generated using the \texttt{genPositiveDefMat} function from the \textsf{R} package \texttt{clusterGeneration}. 

Table~\ref{tab:sim4_pe} gives the true parameter values and the average of the estimates, with standard deviations, returned by MSALD-Bayes and MSALD-EM. The increase in dimension does not appear to affect the overall performance of either approach. Compared to the previous simulation studies, there are many similar trends. For example, we observe that the majority of estimates returned by the MSALD-Bayes are closer to the true values than those returned by the MSALD-EM. In addition, we observe that the MSALD-Bayes is typically more precise than the MSALD-EM. 

\begin{table}[!ht]
\tiny
\centering
\caption{True parameter values and the mean estimates with standard deviations returned by the MSALD-Bayes and MSALD-EM across the 100 simulated data sets generated for simulation 4.}
\label{tab:sim4_pe}
\begin{tabular}{cccc}
\toprule  
\textbf{Parameter} & \textbf{True values} & \textbf{SAL-Bayes} & \textbf{SAL-EM} \\
\midrule  
${\bm{\alpha}}_1$
&
$\left(
\begin{array}{c}
2.30 \\
1.79 \\
0.69 \\
\end{array} \right)
$
&
\textcolor{black}{$
\left(
\begin{array}{c}
2.34 \pm 0.23 \\
1.77 \pm 0.24 \\
0.71 \pm 0.18 \\
\end{array} 
\right)
$}
& 
$
\left(
\begin{array}{c}
2.32 \pm 0.31 \\
1.73 \pm 0.25 \\
0.66 \pm 0.23
\end{array} 
\right)
$
\\
\\
${\bm{\Sigma}}_1$
&
$
\left(
\begin{array}{ccc}
6.29 & -1.32 & 0.68 \\
-1.32 & 7.56 & -0.04 \\
0.68  & -0.04 & 3.89 \\
\end{array} \right)
$
&
\textcolor{black}{$
\left(
\begin{array}{ccc}
6.38 \pm 0.94 & -1.44 \pm 0.59 & 0.65 \pm 0.54 \\
-1.44 \pm 0.59 & 7.64 \pm 1.09 & -0.08 \pm 0.37 \\
0.65 \pm 0.54 & -0.08 \pm 0.37 & 4.05 \pm 0.56 \\
\end{array} \right)
$}
&
$
\left(
\begin{array}{ccc}
6.55 \pm 2.32 & -1.36 \pm 0.59 & 0.55 \pm 2.02 \\
-1.36 \pm 0.59 & 7.62 \pm 1.09 & -0.03 \pm 0.37 \\
0.55 \pm 2.02 & -0.03 \pm 0.37 & 4.12 \pm 1.53 \\
\end{array} \right)
$
\\
\\
${\bm{\mu}}_1$
& 
$
\left(
\begin{array}{c}
0 \\
0 \\
0 \\
\end{array} \right)
$
&
\textcolor{black}{$
\left(
\begin{array}{c}
-0.01 \pm 0.12 \\
-0.01 \pm 0.13 \\
0.00 \pm 0.09
\end{array} 
\right)
$}
&
$
\left(
\begin{array}{c}
-0.00 \pm 0.53 \\
0.04 \pm 0.16 \\
0.07 \pm 0.46
\end{array} 
\right)
$ \\
\\
$\pi_1$ 
&
0.5
& \textcolor{black}{$0.50 \pm 0.02$} 
& $0.50 \pm 0.03$ \\
\midrule
${\bm{\alpha}}_2$
&
$\left(
\begin{array}{c}
-0.60 \\
1.54 \\
3.43 \\
\end{array} \right)
$
&
\textcolor{black}{$
\left(
\begin{array}{c}
-0.61 \pm 0.19 \\
1.54 \pm 0.20 \\
3.43 \pm 0.24
\end{array} 
\right)
$}
& 
$
\left(
\begin{array}{c}
-0.54 \pm 0.52 \\
1.51 \pm 0.22 \\
3.29 \pm 0.54
\end{array} 
\right)
$
\\
\\
${\bm{\Sigma}}_2$
&
$
\left(
\begin{array}{ccc}
4.73 & -1.41 & 0.71 \\
-1.41 & 4.63 & 0.04 \\
0.71 & 0.04 & 1.19 \\
\end{array} \right)
$
&
\textcolor{black}{$
\left(
\begin{array}{ccc}
4.76 \pm 0.56 & -1.49 \pm 0.34 & 0.68 \pm 0.36 \\
-1.49 \pm 0.34 & 4.71 \pm 0.50 & 0.01 \pm 0.34 \\
0.68 \pm 0.36 & 0.01 \pm 0.34 & 1.16 \pm 0.33 \\
\end{array} \right)
$}
&
$
\left(
\begin{array}{ccc}
5.01 \pm 2.41 & -1.53 \pm 0.35 & 0.39 \pm 1.99 \\
-1.53 \pm 0.35 & 4.77 \pm 0.50 & 0.170 \pm 0.44 \\
0.30 \pm 1.99 & 0.17 \pm 0.44 & 1.74 \pm 1.85 \\
\end{array} \right)
$
\\
\\
${\bm{\mu}}_2$
& 
$
\left(
\begin{array}{c}
-4.92 \\
0.24 \\
4.32 \\
\end{array} \right)
$
&
\textcolor{black}{$
\left(
\begin{array}{c}
-4.92 \pm 0.10 \\
0.20 \pm 0.10 \\
4.31 \pm 0.06
\end{array} 
\right)
$}
&
$
\left(
\begin{array}{c}
-4.95 \pm 0.13 \\
0.24 \pm 0.13 \\
4.43 \pm 0.19
\end{array} 
\right)
$ \\
\\
$\pi_2$ 
&
0.5
& \textcolor{black}{$0.50 \pm 0.02$}
& $0.5 \pm 0.03$ \\
\bottomrule
\end{tabular}
\end{table}


\subsection{Classification performance, model selection, and efficiency}\label{sec:allsim_compare}

In the preceding sections, we focused on parameter recovery. In \cite{franczak14}, the MSALD are introduced as a tool for model-based classification in unsupervised and semi-supervised applications. As such, we now turn our attention to comparing the classification performance of the MSALD-Bayes and MSALD-EM. To compare classification performance, we use the ARI. In addition, we also compare the efficacy of using the BIC and ICL for model selection and the run times of the two procedures. 

In Table~\ref{tab:ari_sims}, we provide the average ARI values, with standard deviation, for the best fitting models as selected by the BIC and ICL. Table~\ref{tab:ari_sims} shows that both algorithms return either good or excellent recoveries of the group structure across the four simulations, with the MSALD-Bayes out-performing the MSALD-EM in simulation studies 1, 2, and 4.

\begin{table}[!ht]
\centering
\caption{Average ARI with standard deviation for the best fitting model as selected by either the BIC or ICL for the MSALD-Bayes and MSALD-EM in each simulation study.}
\label{tab:ari_sims}
\begin{tabular}{cccccc}
\toprule  
& \multicolumn{2}{c}{\textbf{MSALD-Bayes}} & 
& \multicolumn{2}{c}{\textbf{MSALD-EM}} \\
\cline{2-3} \cline{5-6}
\textbf{Simulation} & \textbf{BIC} & \textbf{ICL} & & \textbf{BIC} & \textbf{ICL}\\
\midrule
\textbf{1} & \textcolor{black}{1.00 (0.01)} & \textcolor{black}{1.00 (0.01)} & & 0.99 (0.03) & 0.99 (0.03) \\
\textbf{2} & \textcolor{black}{0.97 (0.02)} & \textcolor{black}{0.97 (0.02)} & & 0.92 (0.11) & 0.92 (0.11) \\
\textbf{3} & \textcolor{black}{0.87 (0.03)} & \textcolor{black}{0.87 (0.03)} & & 0.88 (0.03) & 0.88 (0.03) \\
\textbf{4} & \textcolor{black}{1.00 (0.01)} &\textcolor{black}{1.00 (0.05)} & & 0.84 (0.06) & 0.85 (0.04) \\
\bottomrule
\end{tabular}
\end{table}

The noted improvement in classification performance found for the MSALD-Bayes can be partially attributed to model selection. As shown in Table~\ref{tab:modelselect_sims}, which gives the number of times the BIC and ICL selected the correct number of mixture components in simulation studies 1 to 4, we see the BIC and ICL are less reliable for the MSALD-EM in simulations 2 and 4. Notably, the BIC and ICL only select the correct number of components for the MSALD-EM for 41\% of the data sets generated for simulation 2. Across all four simulation studies, we see that the BIC and ICL are much more reliable for the MSALD-Bayes, as both criterion select the correct number of mixture components for at least 97\% of the simulated data sets.

\begin{table}[!ht]
\centering
\caption{Number of times the BIC and ICL selected the correct number of mixture components for the MSALD-Bayes and MSALD-EM in each simulation study.}
\label{tab:modelselect_sims}
\begin{tabular}{ccccccccc}
\toprule  
& & \multicolumn{4}{c}{\textbf{Simulation}} \\
& & \textbf{1} & \textbf{2} & \textbf{3} & \textbf{4} \\ 
\midrule
\multirow{2}{*}{\textbf{MSALD-Bayes}} & \textbf{BIC} & \textcolor{black}{97} & \textcolor{black}{97} & \textcolor{black}{100} & \textcolor{black}{100} \\
& \textbf{ICL} & \textcolor{black}{97} & \textcolor{black}{97} & \textcolor{black}{100} & \textcolor{black}{100}\\
\multirow{2}{*}{\textbf{MSALD-EM}} & \textbf{BIC} & 96 & 41 & 96 & 77 \\
& \textbf{ICL} & 97 & 41 & 97 & 90 \\
\bottomrule
\end{tabular}
\end{table}

Table~\ref{tab:time_sims} gives the elapsed cpu times (in seconds), with standard deviations, for the MSALDs selected by both the BIC and ICL. It is not surprising to observe that the MSALD-Bayes returned higher average elapsed cpu times compared to the MSALD-EM. We see that the elapsed times for the BIC and ICL are the same across all four simulations for the MSALD-Bayes. This represents agreement between the BIC and ICL across all four simulated data sets. Notably, we do not observe the same agreement between the BIC and ICL for the MSALD-EM. This is most likely a result of a higher degree of uncertainty in the classification performance for the selected MSALD-EM.

\begin{table}[!ht]
\centering
\caption{Average elapsed time, in seconds, with standard deviation for the best fitting MSALD models as selected by the BIC and ICL for each simulation study}
\label{tab:time_sims}
\begin{tabular}{cccccc}
\toprule  
& \multicolumn{2}{c}{\textbf{MSALD-Bayes}} & 
& \multicolumn{2}{c}{\textbf{MSALD-EM}} \\
\cline{2-3} \cline{5-6}
\textbf{Simulation} & \textbf{BIC} & \textbf{ICL} & & \textbf{BIC} & \textbf{ICL}\\
\midrule
\textbf{1} & \textcolor{black}{$6276 \pm 4800$} & \textcolor{black}{$6276 \pm 4800$} & & $0.26 \pm 0.49$ & $0.26 \pm 0.49$ \\
\textbf{2} & \textcolor{black}{$7329 \pm 3775$} & \textcolor{black}{$7329 \pm 3775$} & & $1.23 \pm 1.61$ & $1.23 \pm 1.61$ \\
\textbf{3} & \textcolor{black}{$5471 \pm 2008$} & \textcolor{black}{$5471 \pm 2008$} & & $0.82 \pm 3.04$ & $0.76 \pm 2.99$ \\
\textbf{4} & \textcolor{black}{$4569 \pm 1725$} & \textcolor{black}{$4569 \pm 1725$} & & $0.44 \pm 0.50$ & $0.32 \pm 0.27$ \\
\bottomrule
\end{tabular}
\end{table}

\section{Real Data Analysis}\label{sec:real_data}

In this section, we fitted the MSALD to two well-known real data sets using our novel Bayesian parameter estimation scheme and EM-algorithm developed in \cite{franczak14}. We consider the famous Old Faithful geyser data set available in \textsf{R} as \texttt{faithful} and the subset of yeast data set available in the \texttt{MixSAL} package as \texttt{yeast}. The MSALD-Bayes and MSALD-EM are fitted to both data sets for $g = 1,\ldots,3$ groups using the initialization procedure and convergence criterion described at the start of section~\ref{sec:sim_study}. If a 3-component solutions if selected, we fitted an extra component until a solution that is not on the upper-bound is found.

\subsection{Old Faithful feyser data}
\begin{figure}[!ht]
\centering
\includegraphics[width=0.95\linewidth,height=0.375\textheight]{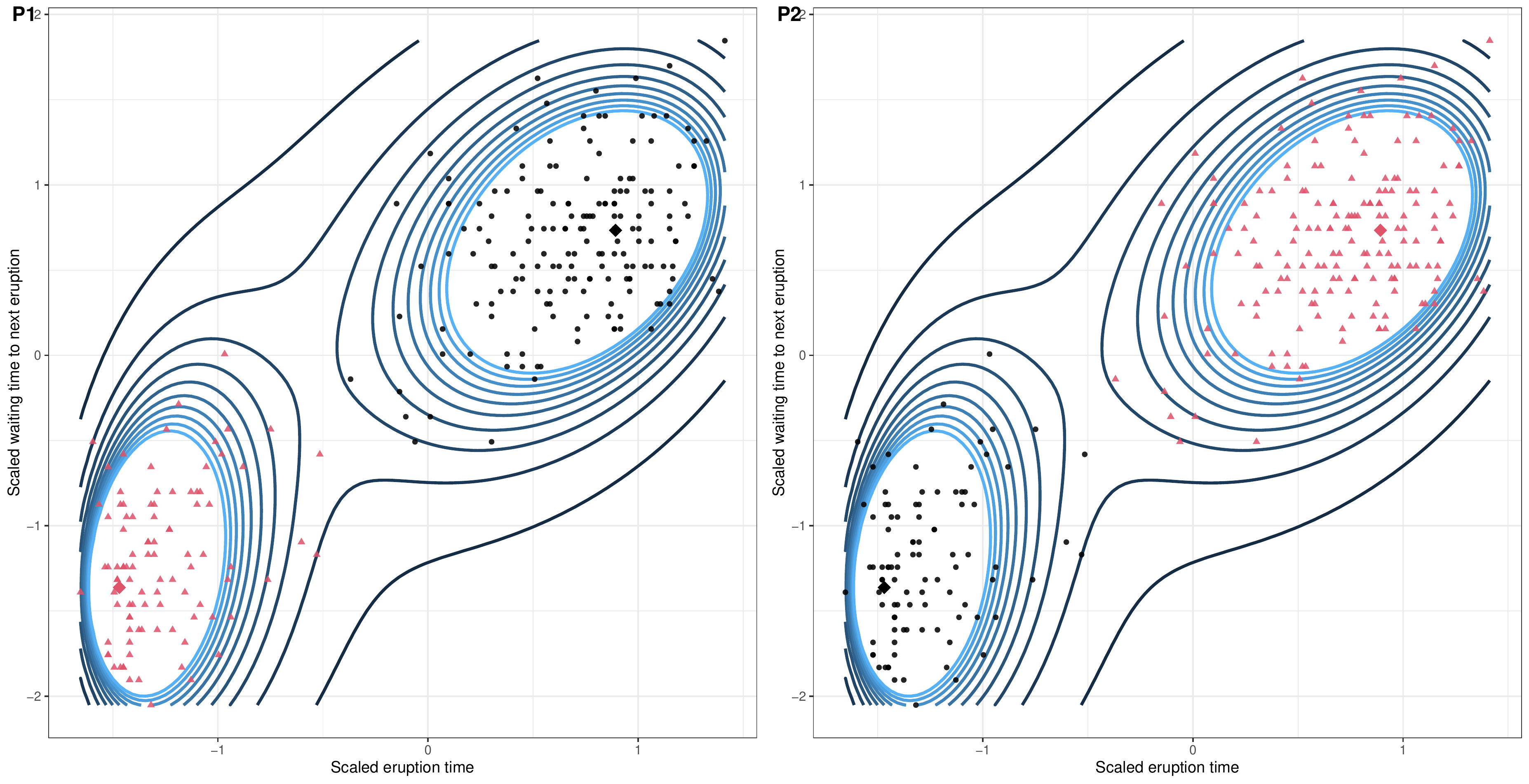}
\caption{Panel 1 (P1) and panel 2 (P2) give the contours and classification results for the best fitting MSALD-Bayes and MSALD-EM solutions chosen by the ICL.}
\label{fig:geyser_contours}
\end{figure}

The Old Faithful geyser data set gives the time between eruptions, in minutes, and the duration of eruptions, in minutes, for 272 eruptions of the Old Faithful geyser in Yellowstone National Park, Wyoming, USA. Figure~\ref{fig:geyser_contours} displays the contour plots with predicted classes coloured based on the best fitting solutions for the MSALD-Bayes and MSALD-EM.

For both the MSALD-Bayes and MSALD-EM, the BIC and ICL select a two-component solution. For the MSALD-Bayes, the best fitting model had $\text{BIC} = -867.08$ and $\text{ICL} = -869.89$. For the MSALD-EM, the best fitting model had $\text{BIC} = -856.52$ and $\text{ICL} = -859.54$. The classification results for the chosen two-component solutions are identical (ARI = 1.00). On both panels 1 and 2, the mean vectors are marked with a solid diamond. It is clear from Figure~\ref{fig:geyser_contours} that the fits are very similar. For both solutions, Figure~\ref{fig:geyser_contours} shows the flexibility of the MSALD and what appear to be very reasonable solutions that accommodates for the natural skewness in the data set. Table~\ref{tab:faithful_pe} gives the parameter estimates for both procedures. The results show that the proposed Bayesian estimate scheme and the EM-algorithm give very similar estimates.

\begin{table}[!ht]
\centering
\caption{Parameter estimates from the best fitting MSALD-Bayes and MSALD-EM for the Old Faithful geyser.}
\label{tab:faithful_pe}
\begin{tabular}{cccccc}
\toprule  
& \multicolumn{2}{c}{\textbf{Component 1}} & & \multicolumn{2}{c}{\textbf{Component 2}} \\
\cline{2-3}\cline{5-6} 
\textbf{Parameter} & \textbf{MSALD-Bayes} & \textbf{MSALD-EM} & & \textbf{MSALD-Bayes} & \textbf{MSALD-EM} \\
\midrule  
${\bm{\alpha}}$
&
$
\left(
\begin{array}{c}
0.20 \\
0.16 \\
\end{array} 
\right)
$
& 
$
\left(
\begin{array}{c}
0.21 \\
0.11 \\
\end{array} 
\right)
$
& & 
$
\left(
\begin{array}{c}
-0.20 \\
-0.08
\end{array} 
\right)
$
& 
$
\left(
\begin{array}{c}
-0.20 \\
-0.12
\end{array} 
\right)
$
\\
\\
${\bm{\Sigma}}$
&
$
\left(\begin{array}{cc}  
0.15 & 0.07 \\
0.07 & 0.25 
\end{array}\right)
$
&
$
\left(\begin{array}{cc}  
0.15 & 0.06 \\
0.06 & 0.25
\end{array}\right)
$
& & 
$
\left(\begin{array}{cc}  
0.02 & 0.01 \\
0.01 & 0.25
\end{array}\right)
$
&
$
\left(\begin{array}{cc}  
0.03 & 0.01 \\
0.01 & 0.25
\end{array}\right)
$
\\\\
${\bm{\mu}}$
&
$
\left(
\begin{array}{c}
0.89 \\
0.73
\end{array} 
\right)
$
&
$
\left(
\begin{array}{c}
0.89 \\
0.77 \\
\end{array} 
\right)
$ 
& & 
$
\left(
\begin{array}{c}
-1.47 \\
-1.36 
\end{array} 
\right)
$
&
$
\left(
\begin{array}{c}
-1.48  \\
-1.32
\end{array} 
\right)
$
\\\\
${\pi}$ 
& 0.65
& 0.64 
& & 0.36
& 0.35
\\
\bottomrule
\end{tabular}
\end{table}

\subsection{Yeast data}
\begin{figure}[!ht]
\centering
\includegraphics[width=0.85\linewidth,height=0.395\textheight]{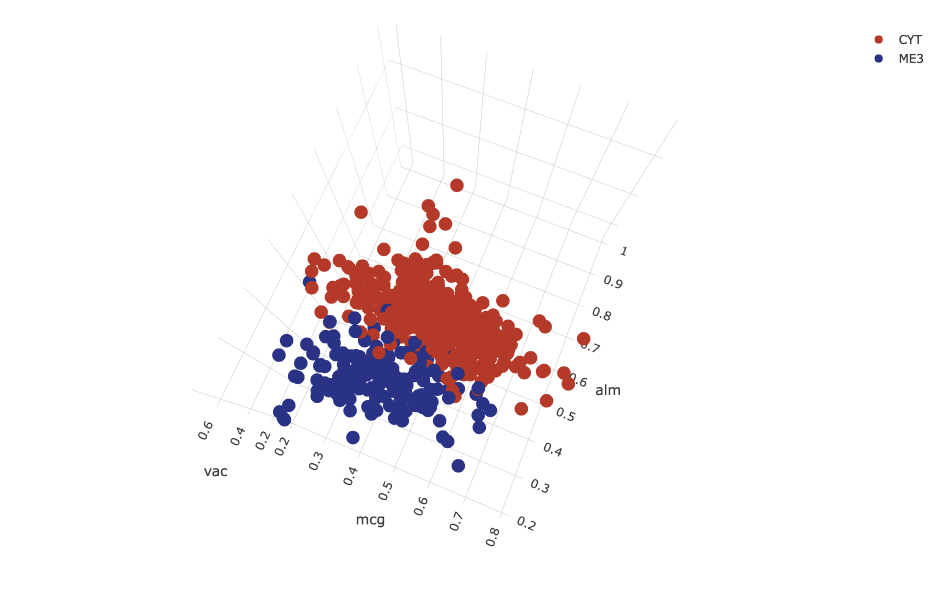}
\caption{A 3D scatterplot showing the CYT and ME3 cellular localization sites of the considered subset of the yeast data set.}
\label{fig:yeast_3d}
\end{figure}

The complete yeast data set is a part of the UCI machine learning repository \citep{Dua19}. \cite{nakai91}, \cite{nakai92} and \cite{horton96} discuss the development of this  data set and classification systems for predicting the cellular localization sites of the proteins it contains. In this analysis, we consider a subset that contains three variables: McGeoch’s method for signal sequence recognition (mcg), the score of the ALOM membrane spanning region prediction program (alm), and the score of discriminant analysis of the amino acid content of vacuolar and extracellular proteins (vac) for 626 proteins. The proteins belong to one of two cellular localization sites: cytosolic or cytoskeletal (CYT) or membrane protein, no N-terminal signal (ME3). One angle of this subset that illustrates the significant amount of overlap between the two localization sites is given in Figure~\ref{fig:yeast_3d}. 

For the MSALD-Bayes procedure, both the BIC ($-5153.35$) and ICL ($-5203.12$) select a 2-component mixture. For the MSALD-EM, the BIC ($-4974.04$) selects a model with 4-components, whereas the ICL ($-5110.00$) selects a 2-component mixture. Since the ICL penalizes for classification uncertainty, this disagreement is attributed to the estimated probabilities of group membership associated with the observations in the overlapping region of the data space. 

\begin{figure}[!ht]
\centering
\includegraphics[width=0.95\linewidth,height=0.5\textheight]{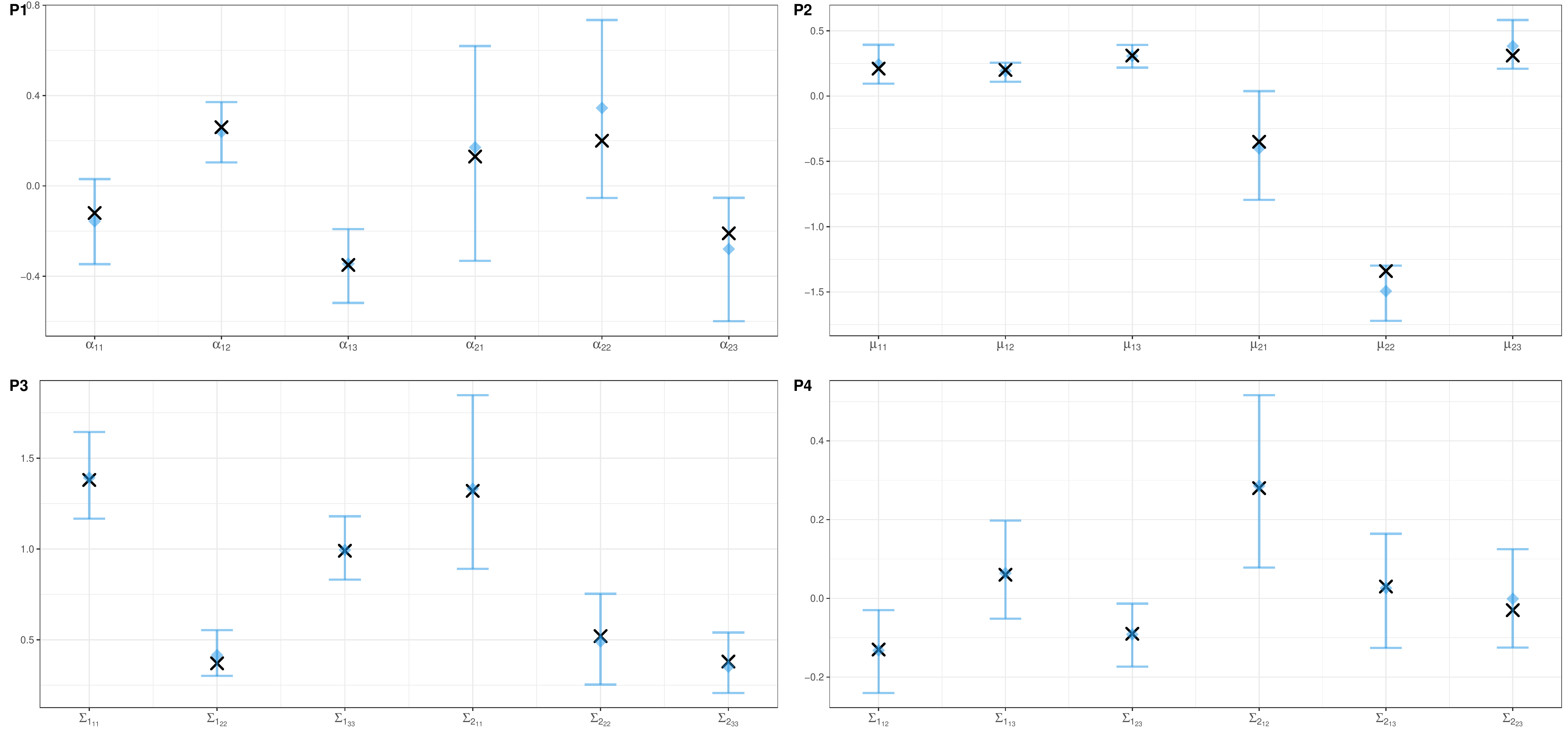}
\caption{Panels 1 through 4 give, respectively, the 95\% percentile intervals in blue from the MSALD-Bayes and the point estimate (black X) from the MSALD-EM for the skewness parameter, location parameter, diagonal elements of the scale matrices, and off-diagonal elements of the scale matrices. The subscripts attached to each parameter designate the component and then the dimension.}
\label{fig:yeast_pe}
\end{figure}

Figure~\ref{fig:yeast_pe} displays the point estimates returned by the MSALD-EM and the 95\% percentile intervals computed using the samples from the posterior distribution of the parameters in the MSALD-Bayes framework. Notably, the point estimates and mid-points of the percentile intervals are very similar for many of the parameters. For the skewness and location parameters, we notice differences in the second dimension of the second component. Figure~\ref{fig:yeast_pe} also displays another advantage of our Bayesian estimation scheme -- the ability to produce an interval estimate for each parameter. Thus, compared to the MSALD-EM, the MSALD-Bayes approach provides a natural framework for assessing the error associated with the parameter estimates.

Despite the many similarities between the point estimates and mid-points of the percentile intervals, we notice that the differences do result in different classification results. Table~\ref{tab:yeast_crosstab} gives the classification tables for the localization sites against the predicted class labels returned by the 2-component MSALD-Bayes and MSALD-EM. Notably, the 2-component MSALD-Bayes (ARI = 0.81) and the 2-component MSALD-EM (ARI = 0.81) return a good recovery of the group structure. However, table~\ref{tab:yeast_crosstab} shows the MSALD-Bayes returned a better hit rate ($0.98$) compared to the MSALD-EM (hit rate = $0.97$).

\begin{table}[!ht]
\centering
\caption{Classification tables of localization sites against the predicted classifications (A, B) for the 2-component MSALD-Bayes and MSALD-EM fitted to the considered subset of the yeast data set.}
\label{tab:yeast_crosstab}
\begin{tabular}{ccccccccc}
\toprule  
& \multicolumn{2}{c}{\textbf{MSALD-Bayes}} & & \multicolumn{2}{c}{\textbf{MSALD-EM}} \\
& A & B & & A & B \\ 
\midrule
CYT & 454 & 9 & & 449 & 14 \\
ME3 & 20 & 143 & & 15 & 148 \\
\bottomrule
\end{tabular}
\end{table}

Interestingly, the $\text{MAP}(z_{ig})$ also lead to slightly different expected values for each group. Table~\ref{tab:yeast_pe} gives the absolute differences between the component-specific sample means and estimated values of component-specific $\E\left[\mathbf{X}\right]=\hat{\vecmu}_g+\hat{\vecalpha}_g$
computed using the midpoints of the 95\% percentile intervals returned by the MSALD-Bayes for each group identified in the 2-component solution. It also provides the absolute differences between the component-specific sample means and estimated values of component-specific $\E\left[\mathbf{X}\right]=\hat{\vecmu}_g+\hat{\vecalpha}_g$ computed using the estimated returned by the MSALD-EM using the 2-component solution identified by the ICL. The results are, again, quite similar with each approach giving estimates that are quite close to the estimated expected values.

\begin{table}[!ht]
\centering
\caption{Absolute differences between the component-specific sample means and estimated values of component-specific $\E\left[\mathbf{X}\right]=\hat{\vecmu}_g + \hat{\vecalpha}_g$ using the estimated parameters by the MSALD-Bayes and the MSALD-EM for each component of the selected best-fitting mixtures.}
\label{tab:yeast_pe}
{
\begin{tabular}{ccccc}
\toprule  
\multicolumn{2}{c}{\textbf{Component 1}} & & \multicolumn{2}{c}{\textbf{Component 2}} \\
\cline{1-2}\cline{4-5} \\
\textbf{MSALD-Bayes} & \textbf{MSALD-EM} & & \textbf{MSALD-Bayes} & \textbf{MSALD-EM} \\
\midrule  
$
\left(
\begin{array}{c}
0.02 \\
0.01 \\
0.00 \\
\end{array} 
\right)
$
& 
$
\left(
\begin{array}{c}
0.01 \\
0.01 \\
0.01 \\
\end{array} 
\right)
$
& & 
$
\left(
\begin{array}{c}
0.10 \\
0.22 \\
0.01 \\
\end{array} 
\right)
$
& 
$
\left(
\begin{array}{c}
0.05\\
0.20 \\
0.05 \\
\end{array} 
\right)
$
\\
\bottomrule
\end{tabular}
}
\end{table}

\section{Discussion}\label{sec:discuss}

Mixtures of shifted asymmetric Laplace (SAL) distributions (MSALDs) were introduced in \cite{franczak14} as an alternative to the mixtures of Gaussian distributions that parameterize skewness, in addition to location and scale. \cite{franczak14} developed an EM algorithm to fit the MSALDs that accounted for the `infinite likelihood problem'. While the MSALDs were shown to give a good classification performance in both simulation and real data applications, an improvement could be made in regards to parameter recovery. In this paper, we propose a novel fully Bayesian parameter estimation scheme for fitting the MSALDs. Our proposed scheme utilizes conjugate priors for the location, scale, and skewness parameters and a Gibbs sampling framework. Furthermore, our proposed approach provides a natural framework to obtain interval estimates as opposed to a traditional EM framework.

Using a variety of simulation studies, we show that the classification performance of the MSALD are quite similar for bi-variate data sets. However, the parameter estimates returned by the proposed Bayesian parameter estimation scheme were usually closer to the true values, on average, with tighter bounds. For the simulated three-dimensional data set, we noted that both the parameter recovery and classification performance was better for the Bayesian scheme. We observed a similar finding in the real data analyses. For the famous Old Faithful geyser data and the considered subset of the yeast data set, both parameter estimation schemes returned similar results, however,  for the considered subset of the Yeast data, the 2-component MSALD-Bayes gave a better hit rate and allows one to directly assess the error associated with parameter estimates.

One drawback of the proposed Bayesian parameter estimation scheme is run-time. Through the considered simulation studies, we show that the proposed scheme can take longer to converge than the EM algorithm. However, while this is not convenient when running repeated simulations, for one data set, one could argue that it is worth the long run time for a more informative result. 

There are a number of directions for future work. One could explore the proposed Bayesian estimate scheme from a strictly computational point of view and focus on reducing the time it takes to fit the MSALD using this approach. Alternatively, the proposed scheme could be utilized to fit parsimonious variations of the MSALD. Another interesting idea is to derive a new Bayesian parameter estimation scheme to fit mixtures of contaminated SAL distributions \citep[MCSALD;][]{morris19}. As noted by the authors, the EM-based parameter estimation scheme developed for the MCSALD must also take into account the `infinite likelihood problem' and therefore, most likely suffers from the same drawbacks illustrated in this paper. 

\section*{Acknowledgements}
This work was supported by a discovery grant from the Natural Sciences and Engineering Research Council of Canada (Franczak, Subedi), a collaboration grant from Simons Foundation (Subedi), and Canada Research Chair program (Subedi). 

\bibliographystyle{chicago}
\bibliography{sal_bayes}

\appendix

\end{document}